\documentclass{ws-procs9x6}
\newcommand{\ergs}{\ifmmode {\rm ergs\,s}^{-1} \else ergs\,s$^{-1}$\fi}
\newcommand{\hb}{H\,$\beta$}

\newcommand{\civ}{C\,{\sc iv}}
\newcommand{\ciii}{C\,{\sc iii}]}
\newcommand{\siiv}{Si\,{\sc iv}}

\newcommand{\heii}{He\,{\sc ii}}
\newcommand{\lam}{$\lambda$}
\newcommand{\mbh}{$M_{\rm BH}$}
\newcommand{\lbol}{$L_{\rm bol}$}
\newcommand{\lol}{$L_{\rm bol}/L_{\rm Edd}$}
\newcommand{\Msol}{\mbox{$M_{\odot}$}}
\newcommand{\lsim}{\stackrel{\scriptscriptstyle <}{\scriptstyle {}_\sim}}
\newcommand{\gsim}{\stackrel{\scriptscriptstyle >}{\scriptstyle {}_\sim}}

\newcommand{\eg}{\mbox{\it e.g.,}\ }
\newcommand{\ie}{\mbox{\it i.e.,}\ }

\begin{document}

\title{Black Hole Masses of High-Redshift Quasars}

\author{M. Vestergaard}

\address{Steward Observatory\\
933 N. Cherry Avenue, \\ 
Tucson, AZ, 85721, USA\\ 
E-mail: mvestergaard@as.arizona.edu}

\maketitle

\abstracts{
Black-hole masses of distant quasars cannot be measured directly,
but can be estimated to within a factor 3 to 5 using scaling 
relationships involving the quasar luminosity and broad-line width.
Why such relationships are reasonable is summarized. The results
of applying scaling relationships to data of quasars at a range
of redshifts ($z \leq 6.3$) are presented.
Luminous quasars typically have masses $\sim 10^9$\Msol{}
even at the highest redshifts. The fact that such massive black
holes appear as early as at $z \approx 6$ indicate that black
holes form very early or build up mass very fast.
}

\section{Introduction}

With our recently acquired ability to measure black-hole masses in
nearby quiescent and active galaxies we are now in a position to start 
addressing
the important issues of the physics of black-hole evolution and the
possible role of black holes in how galaxies evolve. An important first step is to 
establish the typical mass of black holes in AGNs at high redshift
relative to more nearby AGNs. Such a study\cite{paper1} is summarized here.

\section{Mass Estimates}

The most robust method to determine the mass of the central black hole 
in active galaxies is that of reverberation mapping.  However, this method 
is impractical for large samples of luminous and distant quasars
as it takes {\bf many} years to measure quasar masses. The reasons are 
that luminous quasars vary with smaller amplitudes and on longer time 
scales that are further increased by time dilation due 
to their cosmological distances. Mass estimates based on single-epoch data are
therefore very useful, even if less accurate. Of the ``secondary mass
estimators'' used in the literature only a couple are useful at high
redshift, as reviewed by Vestergaard\cite{review}. The ``scaling relations''
used here also appear the most promising at present given their relatively 
lower and readily quantifiable associated uncertainties\cite{review}.

Scaling relations are approximations to the virial mass ($M \propto v^2 R$) 
determination of the reverberation mapping method, where the light travel 
time delay, $\tau = R/c$, between continuum and line variations determine 
the distance $R$ of the line-emitting gas and 
the line width of the variable part of the line profile, the RMS profile,
yields the velocity dispersion $v$ of the same varying gas. 
Instead, scaling relations use 
the empirical radius $-$ luminosity relationship\cite{kaspi}, where 
$R \propto L^{0.7}$, and single-epoch measurements of the line width and the
continuum luminosity to estimate the black-hole mass.  
Vestergaard\cite{calibration} 
calibrated single-epoch mass estimates based on \hb{} and \civ{} line 
widths, respectively, to the more accurate reverberation masses.
The associated statistical uncertainty is a factor $\sim 3$ relative to
the reverberation masses (\ie{} a factor $\sim 5$ on an absolute scale).
However, mass estimates of {\em individual objects} can be in error up to a factor
of 10.
The mass estimates presented below are based on these two relations.

Contrary to the belief of some, scaling relations are reasonable, even those
based on UV spectroscopy for the following reasons. First, contemporaneous 
UV-optical monitoring of
NGC~5548, the best-studied nearby Seyfert, shows that all the broad lines
measured (\siiv \lam 1400, \civ \lam 1549, \heii \lam 1640, \ciii \lam 1909,
\hb \lam 4861, \heii \lam 4686) are consistent with virial motion of the
broad-line region\cite{pw99,pw00}: higher ionization lines are emitted closer
to the central source and have larger Doppler widths. Second, this virial
relationship is seen for all (four) AGNs that can be tested\cite{pw99,pw00,op02}
and so it is fair to assume the relationship is universal, even if the sample
is small. Third, since the virial product ($v^2 \tau$) is constant for each
AGN, the velocity dispersions and the response-lag scale between the emission
lines. Finally, the $R - L$ relation extends also to high redshift
and high luminosity quasars because (1) quasar spectra are very similar\cite{md02} at all
redshifts and luminosities considered here, and (2) the most luminous quasars
have luminosities not much larger ($\lsim$ 1.5\,dex) than the luminosity range
over which the $R - L$ relationship is defined\cite{paper1}.
See Vestergaard\cite{paper1,review,poster} for further details.

\section{Quasar Samples and Data}
In the following, mass estimates are presented for different
samples of quasars spanning the redshift range $0 \lsim z \lsim 6.2$:
the Bright Quasar Survey  (BQS) 
of 87 objects at $z \leq 0.5$, a sample of 114 $1.5 \lsim z \lsim 3.5$ quasars 
almost equally distributed among radio-quiet and radio-loud sources, 
and $\sim$150 $z \approx 4$ quasars from recently published samples which include 
objects and data from the Sloan Digital Sky Survey.
See Vestergaard\cite{paper1} for details.
Bolometric luminosities were estimated using bolometric corrections to
restframe UV luminosities, based on average
spectral energy distributions\cite{elvis94}, 
updated to include a more realistic X-ray energy distribution\cite{elvis02}.

\begin{figure}[t]
\vspace{-0.6cm}
\centerline{\epsfxsize=1.55in\epsfbox{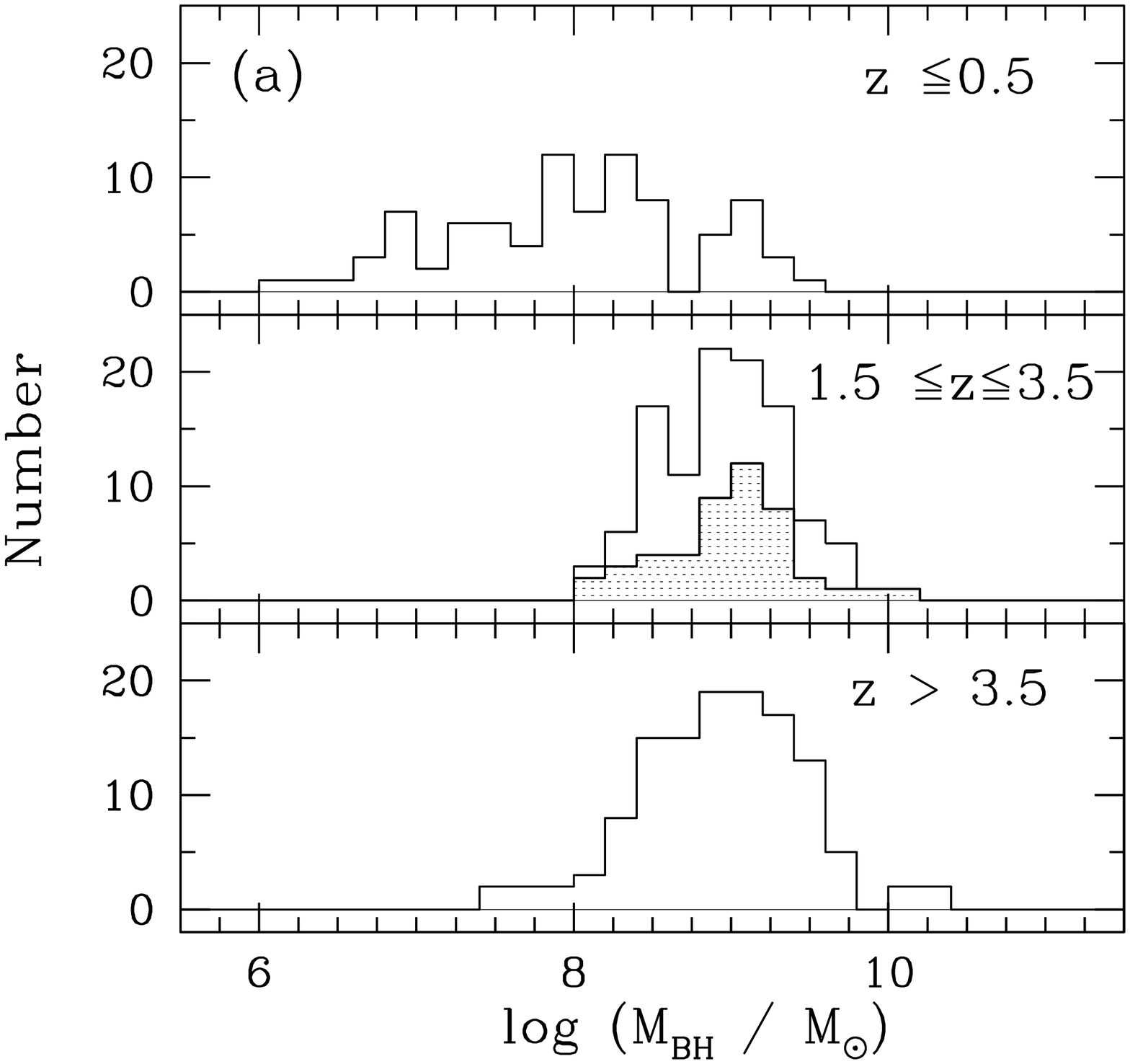}\epsfxsize=1.55in\epsfbox{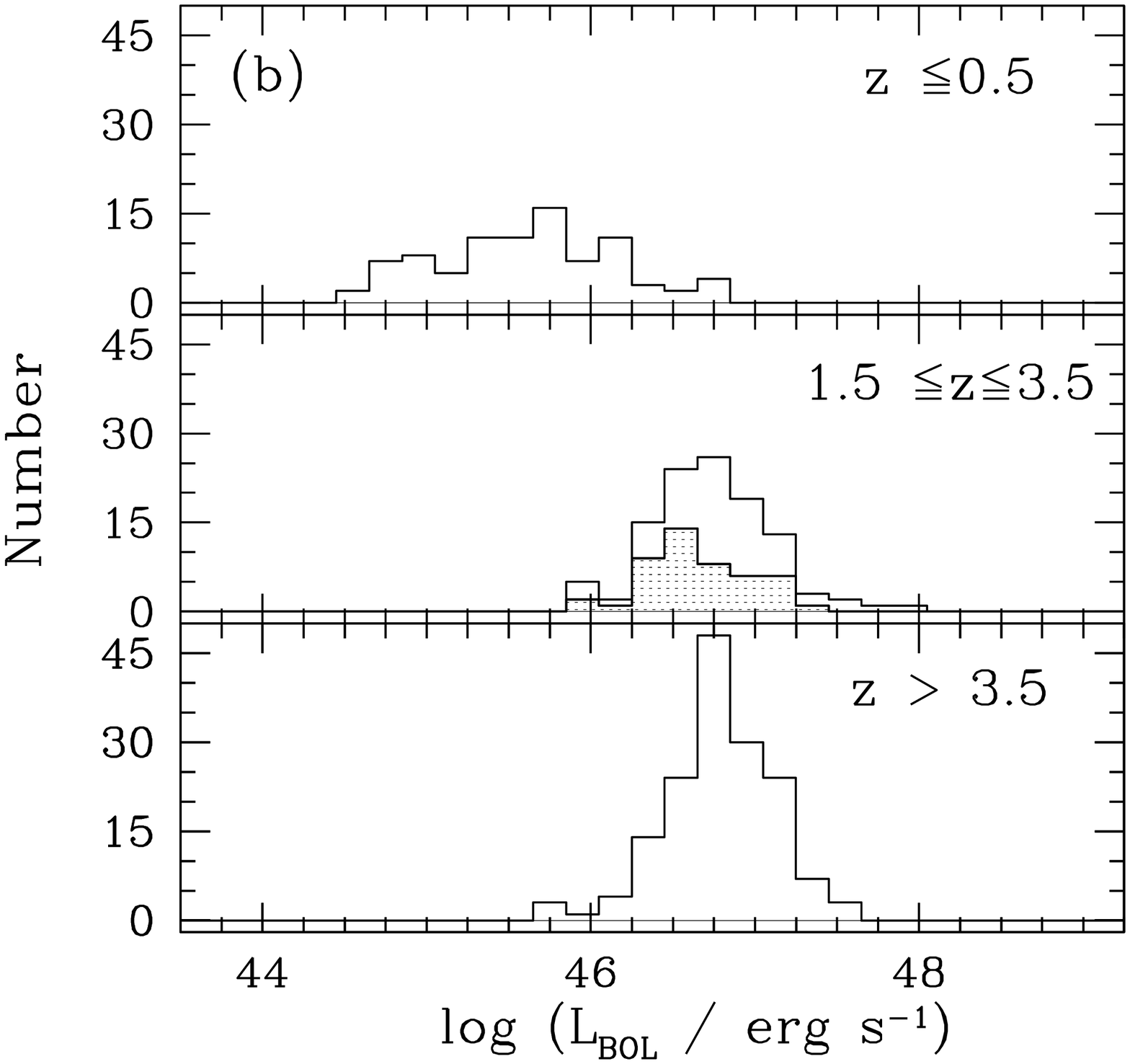}\epsfxsize=1.55in\epsfbox{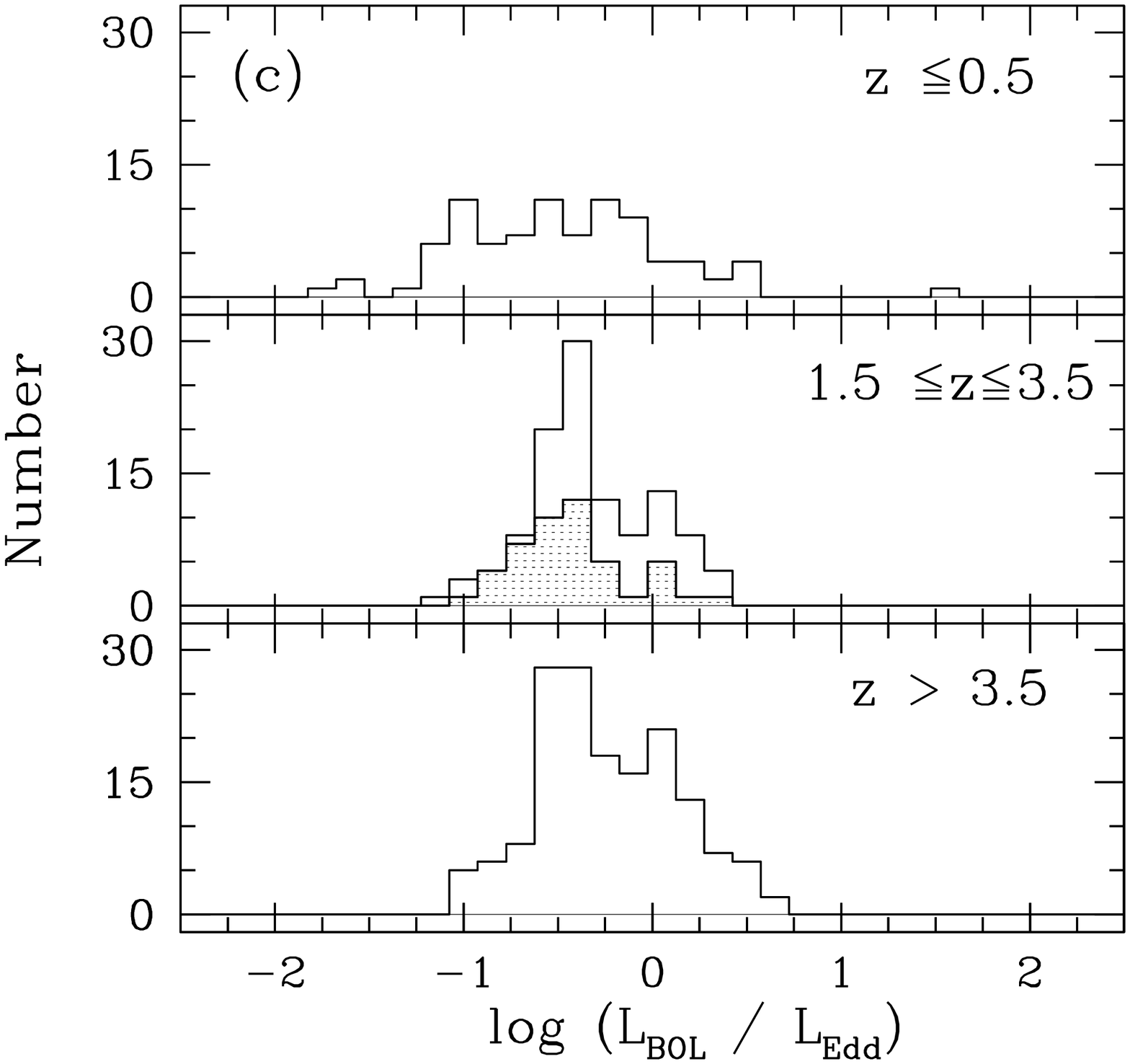}}
\caption{Distributions of estimated (a) black-hole mass, (b) bolometric luminosity,
and (c) Eddington ratio for different redshift bins 
($H_0$ = 75 ${\rm km~ s^{-1} Mpc^{-1}}$, q$_0$ = 0.5, and $\Lambda$ = 0). 
In the middle panel the radio-quiet subset is shown shaded to illustrate that the two 
radio types do {\em not} differ in these parameters as claimed earlier (\eg [12]). 
\label{fig1}}
\end{figure}

\section{Masses of Distant Quasars}

Figure~1 shows that the luminous quasars at $z > 1.5$ are similarly distributed in
black-hole mass \mbh, bolometric luminosity \lbol, and Eddington ratios (\lol) with
an average mass of $10^9$\Msol{} and luminosity of $10^{47}$\ergs. While the lower 
limits in \mbh{} and \lbol{} are due to the sample selection and survey limits, the
data show the important fact that the luminous, distant quasars that we {\em can} 
detect are equally massive as the lower redshift quasars, even beyond the epoch 
($z \gsim 3$) where the comoving quasar space density drops. 
Moreover, there are characteristic, but real, ceilings at \mbh{} $\approx 10^{10}$\Msol{} 
and \lbol{} $\approx 10^{48}$\ergs{}.

\begin{figure}[t]
\vspace{-0.7cm}
\centerline{\epsfxsize=2.25in\epsfbox{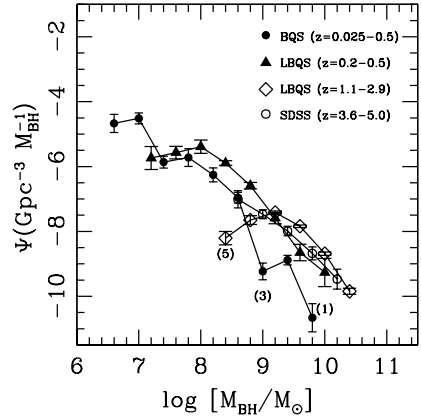}}
\caption{Mass functions of the BQS, LBQS (preliminary), and the 
color-selected SDSS sample 
($H_0$\,=\,50\,${\rm km~ s^{-1} Mpc^{-1}}$, q$_0$\,=\,0.5, $\Lambda$\,=\,0). 
The low-mass turn-down for more distant LBQS 
quasars is an artifact owing to the lower flux-limit of the sample. 
\label{fig2}}
\end{figure}

\section{Mass Functions}

Vestergaard, Osmer, and Fan (in preparation) are currently determining mass 
functions of active black holes at different redshifts. Our first results show 
that the BQS, Large Bright Quasar Survey (LBQS), and color-selected SDSS samples exhibit consistent mass 
functions (Fig.~2; $H_0$ = 50 ${\rm km~ s^{-1} Mpc^{-1}}$, q$_0$ = 0.5, $\Lambda$ = 0). 
The goal is to constrain black-hole growth by combining theoretical models 
(see Steed this volume) with measurements from large data bases.

\section{Conclusions}
The two main conclusions from this work are:
\begin{enumerate}
\item Black-hole masses in nearby AGNs can be measured to within a factor 3.
For more distant AGNs, useful mass {\em estimates} can be obtained to
within a factor of 3 to 5. Even if less accurate, mass estimates are particularly
useful for statistical studies. 

\item Black holes of luminous quasars are very massive ($\sim 10^9$\Msol) even at
the highest redshifts of 4 to 6. The existence of such massive black holes at these
early epochs indicate that they formed very early or very fast.
\end{enumerate}

\end{document}